\documentclass[twocolumn, twocolappendix, tighten]{aastex631}

\def\bpos{B$_{\mathrm{pos}}$}

\begin{document}

\title{Magnetic Fields in Massive Star-forming Regions (MagMaR): Unveiling an Hourglass Magnetic Field in G333.46$-$0.16 using ALMA}

\correspondingauthor{Piyali Saha}
\email{s.piyali16@gmail.com/piyali.saha@nao.ac.jp}

\author[0000-0002-0028-1354]{Piyali Saha}
\affiliation{National Astronomical Observatory of Japan, National Institutes of Natural Sciences, 2-21-1 Osawa, Mitaka, Tokyo 181-8588, Japan}

\author[0000-0002-7125-7685]{Patricio Sanhueza}
\affiliation{National Astronomical Observatory of Japan, National Institutes of Natural Sciences, 2-21-1 Osawa, Mitaka, Tokyo 181-8588, Japan}
\affiliation{Astronomical Science Program, The Graduate University for Advanced Studies, SOKENDAI, 2-21-1 Osawa, Mitaka, Tokyo 181-8588, Japan}

\author[0000-0003-2303-0096]{Marco Padovani}
\affiliation{INAF-Osservatorio Astrofisico di Arcetri, Largo E. Fermi 5, 50125 Firenze, Italy}

\author[0000-0002-3829-5591]{Josep M. Girart}
\affiliation{Institut de Ci\'encies de l’Espai (ICE-CSIC), Campus UAB, Can Magrans S/N, E-08193 Cerdanyola del Vall\'es, Catalonia, Spain}
\affiliation{Institut d’Estudis Espacials de Catalunya (IEEC), Esteve Terradas 1, PMT-UPC, E-08860 Castelldefels, Catalonia, Spain}

\author[0000-0002-3583-780X]{Paulo Cortes}
\affiliation{Joint ALMA Observatory, Alonso de C\'ordova 3107, Vitacura, Santiago, Chile}
\affiliation{National Radio Astronomy Observatory, 520 Edgemont Road, Charlottesville, VA 22903, USA}

\author[0000-0002-6752-6061]{Kaho Morii}
\affiliation{Department of Astronomy, Graduate School of Science, The University of Tokyo, 7-3-1 Hongo, Bunkyo-ku, Tokyo 113-0033, Japan}
\affiliation{National Astronomical Observatory of Japan, National Institutes of Natural Sciences, 2-21-1 Osawa, Mitaka, Tokyo 181-8588, Japan}

\author[0000-0002-4774-2998]{Junhao Liu}
\affiliation{National Astronomical Observatory of Japan, National Institutes of Natural Sciences, 2-21-1 Osawa, Mitaka, Tokyo 181-8588, Japan}

\author[0000-0002-3078-9482]{\'A. S\'anchez-Monge}
\affiliation{Institut de Ci\'encies de l’Espai (ICE-CSIC), Campus UAB, Can Magrans S/N, E-08193 Cerdanyola del Vall\'es, Catalonia, Spain}
\affiliation{Institut d’Estudis Espacials de Catalunya (IEEC), Esteve Terradas 1, PMT-UPC, E-08860 Castelldefels, Catalonia, Spain}

\author[0000-0001-7706-6049]{Daniele Galli}
\affiliation{INAF-Osservatorio Astrofisico di Arcetri, Largo E. Fermi 5, 50125 Firenze, Italy}

\author[0000-0003-0855-350X]{Shantanu Basu}
\affiliation{Department of Physics and Astronomy, University of Western Ontario, London, ON N6A 3K7, Canada}

\author[0000-0003-2777-5861]{Patrick M. Koch}
\affiliation{Academia Sinica, Institute of Astronomy and Astrophysics, Taipei, Taiwan}

\author[0000-0003-3315-5626]{Maria T.\ Beltr\'an}
\affiliation{INAF-Osservatorio Astrofisico di Arcetri, Largo E. Fermi 5, I-50125 Firenze, Italy}

\author[0000-0003-1275-5251]{Shanghuo Li}
\affiliation{Max Planck Institute for Astronomy, K\"onigstuhl 17, D-69117 Heidelberg, Germany}

\author[0000-0002-1700-090X]{Henrik Beuther}
\affiliation{Max Planck Institute for Astronomy, K\"onigstuhl 17, D-69117 Heidelberg, Germany}

\author[0000-0003-3017-4418]{Ian W. Stephens} 
\affiliation{Department of Earth, Environment, and Physics, Worcester State University, Worcester, MA 01602, USA}

\author[0000-0001-5431-2294]{Fumitaka Nakamura}
\affiliation{National Astronomical Observatory of Japan, National Institutes of Natural Sciences, 2-21-1 Osawa, Mitaka, Tokyo 181-8588, Japan}
\affiliation{Department of Astronomy, The University of Tokyo, Hongo, Tokyo 113-0033, Japan}
\affiliation{The Graduate University for Advanced Studies (SOKENDAI), 2-21-1 Osawa, Mitaka, Tokyo 181-0015, Japan}

\author[0000-0003-2384-6589]{Qizhou Zhang}
\affiliation{Center for Astrophysics $\vert$ Harvard \& Smithsonian, 60 Garden Street, Cambridge, MA, 02138, USA}

\author[0000-0001-9822-7817]{Wenyu Jiao}
\affiliation{Department of Astronomy, School of Physics, Peking
University, Beijing, 100871, People's Republic of China}
\affiliation{Kavli Institute for Astronomy and Astrophysics, Peking University, Haidian District, Beijing 100871, People’s Republic of China}

\author[0000-0001-5811-0454]{M. Fern\'andez-L\'opez}
\affiliation{Instituto Argentino de Radioastronom\'\i a (CCT-La Plata, CONICET; UNLP; CICPBA), C.C. No. 5, 1894, Villa Elisa, Buenos Aires, Argentina}

\author[0000-0001-7866-2686]{Jihye Hwang}
\affiliation{Korea Astronomy and Space Science Institute, 776 Daedeokdae-ro, Yuseong-gu, Daejeon 34055, Republic of Korea}

\author[0000-0003-0014-1527]{Eun Jung Chung}
\affiliation{Korea Astronomy and Space Science Institute, 776 Daedeokdae-ro, Yuseong-gu, Daejeon 34055, Republic of Korea}

\author[0000-0002-8557-3582]{Kate Pattle}
\affiliation{Department of Physics and Astronomy, University College London, Gower Street, London WC1E 6BT, United Kingdom}

\author[0000-0003-2343-7937]{Luis A. Zapata}
\affil{Instituto de Radioastronom\'\i a y Astrof\'\i sica, Universidad Nacional Aut\'onoma de M\'exico, P.O. Box 3-72, 58090, Morelia, Michoac\'an, M\'exico}

\author[0000-0001-5950-1932]{Fengwei Xu}
\affiliation{I. Physikalisches Institut, Universität zu Köln, Zülpicher Str. 77, D-50937 Köln, Germany}
\affiliation{Kavli Institute for Astronomy and Astrophysics, Peking University, Beijing 100871, People's Republic of China}
\affiliation{Department of Astronomy, School of Physics, Peking University, Beijing, 100871, People's Republic of China}

\author[0000-0002-8250-6827]{Fernando A. Olguin}
\affiliation{Institute of Astronomy, National Tsing Hua University, Hsinchu 30013, Taiwan}

\author[0000-0001-7379-6263]{Ji-hyun Kang}
\affiliation{Korea Astronomy and Space Science Institute, 776 Daedeok-daero, Yuseong, Daejeon 34055, Republic of Korea}

\author[0000-0001-5996-3600]{Janik Karoly}
\affiliation{Department of Physics and Astronomy, University College London, WC1E 6BT London, UK}

\author[0000-0003-1964-970X]{Chi-Yan Law}
\affiliation{INAF-Osservatorio Astrofisico di Arcetri, Largo E. Fermi 5, I-50125 Firenze, Italy}

\author[0000-0002-6668-974X]{Jia-Wei Wang}
\affiliation{Institute of Astronomy and Department of Physics, National Tsing Hua University, Hsinchu 30013, Taiwan}

\author[0000-0002-6018-1371]{Timea Csengeri}
\affiliation{Laboratoire d'astrophysique de Bordeaux, Univ. Bordeaux, CNRS, B18N, all\'ee Geoffroy Saint-Hilaire, 33615 Pessac, France}

\author[0000-0003-2619-9305]{Xing Lu}
\affiliation{Shanghai Astronomical Observatory, Chinese Academy of Sciences, 80 Nandan Road, Shanghai 200030, People's Republic of China}

\author[0000-0002-8691-4588]{Yu Cheng}
\affiliation{National Astronomical Observatory of Japan, National Institutes of Natural Sciences, 2-21-1 Osawa, Mitaka, Tokyo 181-8588, Japan}

\author[0000-0002-1229-0426]{Jongsoo Kim}
\affiliation{Korea Astronomy and Space Science Institute (KASI), 776 Daedeokdae-ro, Yuseong-gu, Daejeon 34055, Republic of Korea}

\author[0000-0002-7497-2713]{Spandan Choudhury}
\affiliation{Korea Astronomy and Space Science Institute, 776 Daedeok-daero Yuseong-gu, Daejeon 34055, Republic of Korea}

\author[0000-0002-9774-1846]{Huei-Ru Vivien Chen}
\affiliation{Institute of Astronomy and Department of Physics,
National Tsing Hua University, Hsinchu 300044, Taiwan}

\author[0000-0002-8975-7573]{Charles L. H. Hull}
\affiliation{National Astronomical Observatory of Japan, National Institutes of Natural Sciences, 2-21-1 Osawa, Mitaka, Tokyo 181-8588, Japan}



\begin{abstract}
\noindent
The contribution of the magnetic field to the formation of high-mass stars is poorly understood. We report the high-angular resolution ($\sim0.3^{\prime\prime}$, 870 au) map of the magnetic field projected on the plane of the sky (\bpos) towards the high-mass star forming region G333.46$-$0.16 (G333), obtained with the Atacama Large Millimeter/submillimeter Array (ALMA) at 1.2 mm as part of the Magnetic Fields in Massive Star-forming Regions (MagMaR) survey. The \bpos ~morphology found in this region is consistent with a canonical ``hourglass'' which suggest a dynamically important field. This region is fragmented into two protostars separated by $\sim1740$ au. Interestingly, by analysing H$^{13}$CO$^{+}$ ($J=3-2$) line emission, we find no velocity gradient over the extend of the continuum which is consistent with a strong field. We model the \bpos, obtaining a marginally supercritical mass-to-flux ratio of 1.43, suggesting a initially strongly magnetized environment. Based on the Davis–Chandrasekhar–Fermi method, the magnetic field strength towards G333 is estimated to be $5.7$ mG. 
The absence of strong rotation and outflows towards the central region of G333 suggests strong magnetic braking, consistent with a highly magnetized environment. Our study shows that despite being a strong regulator, the magnetic energy fails to prevent the process of fragmentation, as revealed by the formation of the two protostars in the central region. 

\end{abstract}



\section{Introduction} \label{sec:intro}

Star formation is a complex process controlled by several factors, among which gravity, turbulence, and magnetic field play a key role. 
Magnetic fields are believed to oppose the gravitational contraction and fragmentation of dense cores, thus delaying the formation of protostars \citep[e.g.,][]{1999ASIC..540..305M,2007ARA&A..45..565M,2017NatAs...1E.158L,2021ApJ...912..159P,2022ApJ...941...51H}. However, magnetic fields can also help to channel cloud material towards overdense regions, thus acting as a catalyst in the formation of protostars \citep[e.g.,][]{2017A&A...607A...2S,2019FrASS...6....5H}. 

In a strongly magnetic environment, the morphology of the magnetic field remains preserved up to scales of dense molecular cores of $0.01-0.1$ pc 
\citep[e.g.,][]{2014ApJ...794L..18Q,2015Natur.520..518L,2021ApJ...923..204C}. Another possible scenario shows the dragging of the frozen-in magnetic field along with the gas material towards the dense core by the gravitational collapse. This phenomenon creates a pinching effect in the magnetic field lines, leading to an ``hourglass"-like appearance 
\citep[e.g., ][]{2006Sci...313..812G,2009Sci...324.1408G,2009ApJ...707..921R,2009ApJ...700..251T,2013ApJ...769L..15S,2014ApJS..213...13H,2014ApJ...794L..18Q,2018MNRAS.477.2760M,2018ApJ...855...39K,2019A&A...630A..54B,2019ApJ...879...25K,2021ApJ...923..204C,2024ApJ...963L..31H}. This specific structure may be partially due to projection and line-of-sight integration effects that lead to the observed dust polarization. It is of utmost importance to characterize hourglass patterns when observed, in order to better understand the initial conditions of star formation.




Despite being a significant regulator of star formation, the role of the magnetic field during the birth of massive stars is still poorly understood. 
To make progress, a survey Magnetic fields in Massive star-forming Regions (MagMaR) has been carried out. In MagMaR, a total of 30 high-mass star-forming regions were observed at 1.2 mm with the Atacama Large Millimeter/ submillimeter Array (ALMA). 
A few detailed characteristics of some targets, such as G5.89$-$0.39, IRAS 18089$-$1732, and NGC 6334I(N), 
have been discussed by \cite{2021ApJ...913...29F}, \cite{2021ApJ...915L..10S}, and \cite{2021ApJ...923..204C}, respectively. 

Out of the 30 targets, the magnetic field projected on the plane of the sky (B$_\mathrm{POS}$) towards the high-mass star-forming region G333.46$-$0.16 (hereafter, G333) shows the most extended hourglass morphology on a few 1000 au scale \citep[assuming a distance of 2.9 kpc;][]{2019A&A...631A..72L}. This target was studied as a part of the ATLASGAL survey of massive clumps 
by \cite{2014A&A...565A..75C} and \cite{2019A&A...631A..72L}. 
The bolometric luminosity of this target is of 4.4$\times$10$^{3}$ L$_{\odot}$ \citep{2019A&A...631A..72L}. Based on its spectral energy distribution, \cite{2019A&A...631A..72L} estimated a dust temperature and mass of 25.2 K and 282 M$_{\odot}$, respectively for this region. Here, we aim to assess magnetic properties and to investigate the importance of the magnetic field, gravity and turbulence in the star-forming region G333.


\begin{figure*}[ht!]
\includegraphics[height=9.5cm, width=\textwidth]{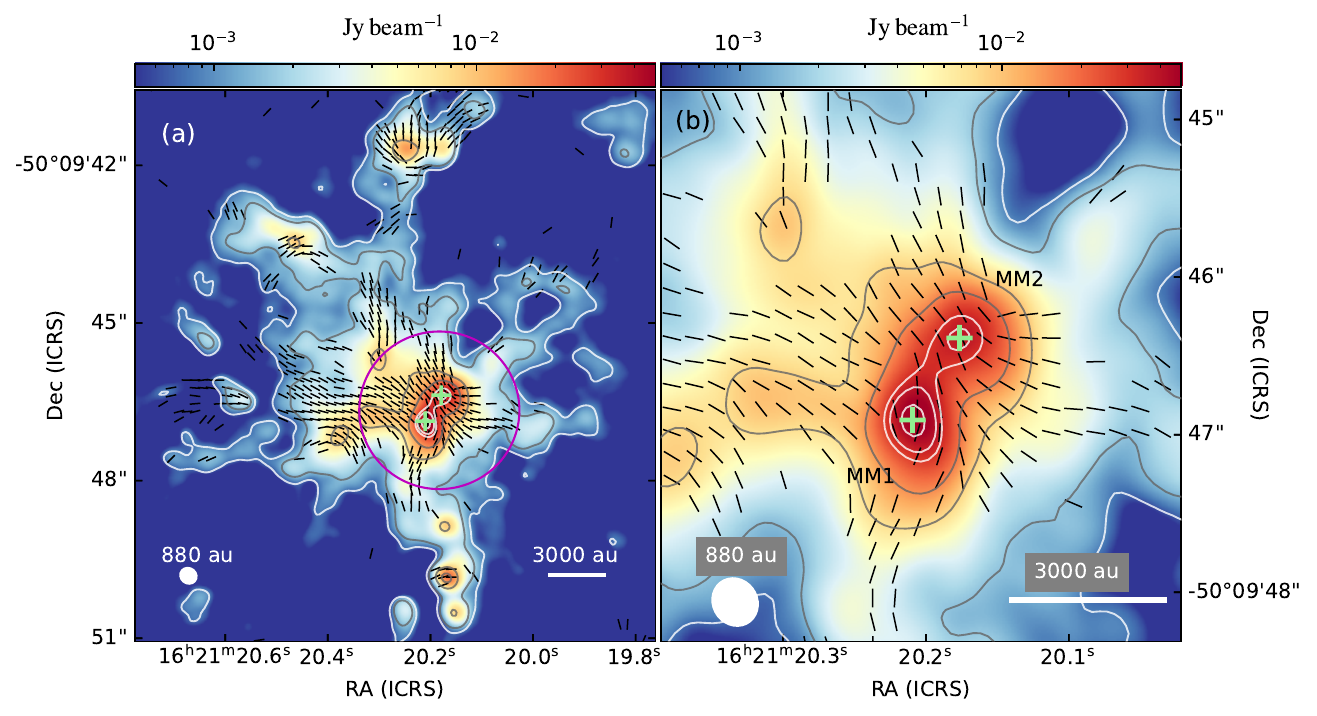}
\caption{\textbf{(a)} The B$_\mathrm{POS}$ geometry (black line segments) towards G333, obtained after rotating the polarization segments by 90$^{\circ}$, overplotted on ALMA 1.2 mm dust continuum emission. The B$_\mathrm{POS}$ segments have arbitrary length and plotted above the 3$\sigma$ ($\sigma=29\mu$Jy beam$^{-1}$ for Stokes $Q$ and $U$) level. Contours correspond to dust continuum emission 5, 10, 50, 100, 190, 230, 300 and 340 times $\sigma$ ($=160 \mu$Jy beam$^{-1}$). The positions of MM1 and MM2 are marked as `+' symbols. The circle in magenta represents the area of analysis. The scale bar and the beamsize of 880 au are displayed on the bottom right and bottom left sides, respectively. The line segments are drawn following the Nyquist sampling (every three pixels each of size $0.05^{\prime\prime}$). \textbf{(b)} Magnified view of the area of analysis. Symbols are the same as in panel \textbf{(a)}.}
\label{fig:bpos}
\end{figure*}

\section{Observations} \label{sec:obs}

ALMA polarimetric observations towards G333 were carried out on 2018 September 27 using ALMA Band 6 (1.2 mm) as a part of the MagMaR project (Project ID: 2017.1.00101.S and 2018.1.00105.S; PI: P. Sanhueza). The maximum recoverable scale (MRS) was $\sim4.5^{\prime\prime}$. 
During the observing runs, J1650-5044 was used as phase calibrator while J1427-4206 was used as a calibrator for bandpass, flux, and polarization. Details of the observational setup are discussed by \cite{2021ApJ...913...29F}, \cite{2021ApJ...915L..10S}, and \cite{2021ApJ...923..204C}.

Linearly polarized dust continuum emission is identified within the the inner one-third area of the primary beam (24$^{\prime\prime}$). 
Bright emission lines have been eliminated from the Stokes $I$ continuum following the method elaborated by \cite{2021ApJ...909..199O}. We used CASA 6.5.2 to perform self-calibration and imaging. Self-calibration of the Stokes $I$ was performed with three iterations in phase with a final solution interval of 10 seconds. The self-calibration solutions were applied to the respective spectral cubes. The imaging of each Stokes parameter was performed separately employing the CASA task \textit{tclean} using Briggs weighting with a robust parameter of 1. The resulting Stokes $I$ image has an angular resolution of 0.318$^{\prime\prime}\times0.290^{\prime\prime}$ with a position angle of 49.4$^{\circ}$ (922$\times$841 au). The sensitivities of the images are 160 $\mu$Jy beam$^{-1}$ for the final Stokes $I$, and 29 $\mu$Jy beam$^{-1}$ for the final Stokes $Q$, and $U$. The source is bright in continuum Stokes $I$ emission making the image dynamic range limited. However, the polarized emission (Stokes $Q$ and $U$) is significantly weaker compared to the continuum Stokes $I$ and therefore, it is unlikely to be dynamic range limited, making them possible to reach the thermal noise. The debiased linear polarized intensity, polarization fraction and polarization angle images were constructed following \cite{2006PASP..118.1340V}.

The image of H$^{13}$CO$^{+} (J=3-2)$ line emission, also included in the spectral setup, was obtained by the automatic masking method \textit{yclean} \citep{2018ApJ...861...14C}. The CASA task \textit{tclean} was performed using Briggs weighting with a robust parameter of 1, leading to a noise level of 2.9 mJy beam$^{-1}$ (0.61 K) per 0.56 km s$^{-1}$ channel.  

\section{Results} \label{sec:res}

The ALMA 1.2 mm continuum emission with $\sim900$ au spatial resolution is shown in Figure \ref{fig:bpos} a and reveals details of the internal structure of G333. The dust continuum emission shows a flattened structure that is elongated in the northwest-southeast direction. The center of this flattened structure is fragmented into two more condensations with a separation of $\sim1740$ au along the major axis of the elongation. Both the condensations show presence of hot molecular core emission lines \citep{2023ApJ...950...57T}, suggesting that they are potential protostars. The peak flux at the position of the brightest component, located towards the south-east (MM1), is $53.5$ mJy beam$^{-1}$, while the fainter one located towards the north-west (MM2) has a peak flux of $38.8$ mJy beam$^{-1}$. 

The direction of the \bpos~ is inferred by assuming the dust grains are aligned with respect to the magnetic fields \citep[i.e., rotating the polarization segments by $90^{\circ}$;][]{1982MNRAS.200.1169C,1984ApJ...284L..51H,1988QJRAS..29..327H,2000ASPC..215...69L,2015ARA&A..53..501A}. 
The polarized emission detected in G333 suggests an hourglass-like geometry of the magnetic field aligned with the symmetry axis of the hourglass almost parallel to the minor axis of the flattened envelope. For our analysis, we focus on the circular area as marked in Figure \ref{fig:bpos} (a), that harbours the hourglass magnetic field. We chose this area to avoid the additional distortion of \bpos ~produced by other surrounding cores. The circle is centered at 
$\alpha=16:21:20.183$ and $\delta=-50:09:46.662$, with a radius of $1.5^{\prime\prime}$ (4350 au). As the diameter of the area of analysis ($3^{\prime\prime}$) is less than the MRS, the extended emission should not be significantly affected by filtering.

\begin{figure}
\includegraphics[width=9.5cm, height=13cm]{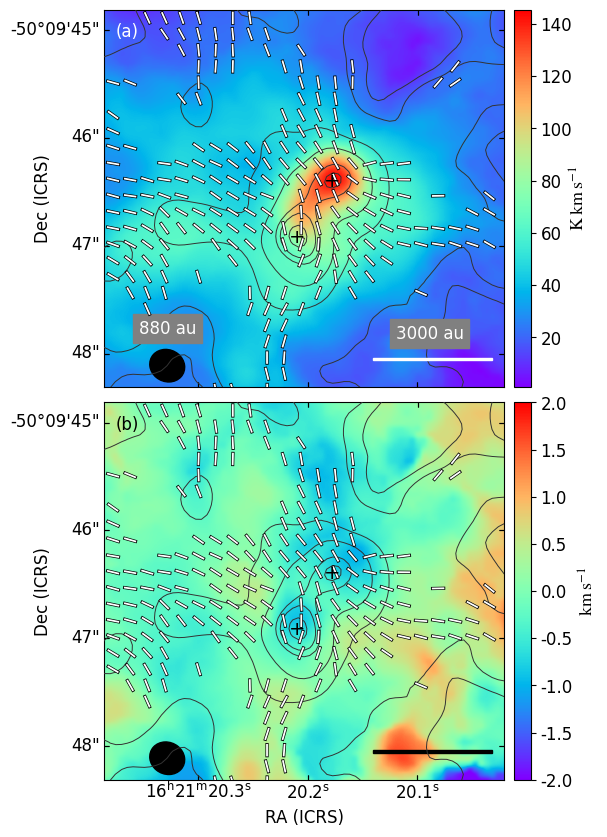}
\caption{\textbf{(a)} The moment 0 or integrated intensity map of H$^{13}$CO$^{+}$ line emission. \textbf{(b)} The moment 1 or intensity weighted velocity map of the same. The white line segments show the B$_\mathrm{POS}$ geometry and the contours outline the dust continuum emission (same as Figure \ref{fig:bpos}). The positions of MM1 and MM2 are marked as `+' symbols. The scale bar and the spatial resolution are displayed on the bottom right and bottom left sides, respectively, in all panels.} 
\label{fig:mom_maps}
\end{figure}

Being a cold dense gas tracer (upper energy level, $E_{u}$ of 25 K), the spatial distribution of H$^{13}$CO$^{+} (J=3-2)$ line emission towards G333 appears to be similar to that of the dust continuum emission (see Figure \ref{fig:mom_maps} a). 
Figure \ref{fig:mom_maps} (b) shows the H$^{13}$CO$^{+}$ distribution of the intensity-weighted velocity structure (moment 1 map) towards G333 with respect to the systemic velocity ($v_\mathrm{lsr}\sim-43$ km s$^{-1}$). No clear signature of large-scale velocity gradient is detected with a spectral resolution of 0.56 km s$^{-1}$, suggesting a quiescent environment at $\sim10000$ au scale. However, the velocity field over MM1 and MM2 is relatively more blueshifted than their immediate vicinities, which could be a sign of infall as suggested by \cite{2019A&A...626A..84E} and \cite{2021ApJ...909..199O}. 


\section{Discussion} \label{sec:dis}

\subsection{{Dust continuum emission and velocity dispersion}} \label{subsec:mass}

The flux density enclosed in the circular region is 493 mJy. 
The temperature of the region marked with a circle as in Figure \ref{fig:bpos}, is determined using the CH$_{3}$CN ($J=14-13$) rotational transitions. We detected primarily 4 brightest K components (K = 0, 1, 2, 3) of the transition towards this area. We avoided the region very close to MM1 and MM2 to refrain the line contamination by radiation from MM1 and MM2. In Figure \ref{fig:ch3cn_temp}, we present the average brightness distribution of the CH$_{3}$CN ($J=14-13$) transitions in black. The blue dashed line represents the fitted spectrum that is obtained by fitting the observed spectrum with XCLASS \citep{2017A&A...598A...7M} resulting in a temperature of 50 K. 
Using this temperature, and assuming optically thin dust emission and a spherical geometry, the total gas mass is estimated using:
\begin{equation}
    M=\Lambda\frac{F_{\nu}D^{2}}{\kappa_{\nu}B_{\nu}(T)}
\end{equation}
\noindent
where $\Lambda$ is the gas-to-dust mass ratio, $F_{\nu}$ is the flux density of the source, $D$ is the distance to the target, $\kappa_{\nu}$ is the dust opacity per gram of dust, and $B_{\nu}$ is the Planck function with a dust temperature T. The gas-to-dust ratio is assumed to be 100:1. The dust opacity is assumed as 1.03 cm$^{2}$g$^{-1}$ \cite[interpolated to 1.2 mm;][]{1994A&A...291..943O}. 
Using a dust temperature of 50 K, the total mass is estimated as 23 M$_{\odot}$.

We estimate the mass density $\rho$ as $3.9\times10^{-17}$ gm cm$^{-3}$, using a volume ($V=\frac{4}{3}\pi R^{3}$, where, $R=1.5^{\prime\prime}$). The number density $n(\rm H_{2})$ is $\rho/\mu_{\rm H_{2}}m_{\rm H}$, where $\mu_{\rm H_{2}}(=2.86)$ is the mean molecular weight per hydrogen molecule \citep{2013MNRAS.432.1424K} and $m_{\rm H}$ is the atomic mass of hydrogen. The $n(\rm H_{2})$ enclosed in the area of analysis is estimated as $8.4\times10^{6}$ cm$^{-3}$.




Using the \textit{dendogram} technique \citep{2008ApJ...679.1338R} adopted in the \textit{astrodendro} Python package \footnote{\url{http://www.dendrograms.org/}}, the flux densities of MM1 and MM2 are estimated as 59.5 mJy and 25.8 mJy, respectively. Assuming the similar temperature, MM1 is 2.3 times more massive than MM2. Table \ref{tab:dedro} lists the \textit{dendogram} results for both protostars. 

Considering a bolometric luminosity of $4.4\times10^{3}$ L$_{\odot}$ \citep{2019A&A...631A..72L}, and assuming that a single main-sequence star is responsible for the bolometric luminosity, the stellar mass would then be $\sim6$ M$_{\odot}$ \citep{2011ApJ...730L..33M}. However, there are two protostars at the center. At larger scale, based on the ALMA 7m array observations with an angular resolution of $3.7^{\prime\prime}$ \citep{2017A&A...600L..10C}, the mass of the structure containing MM1 and MM2 is determined as 120 M$_{\odot}$, suggesting that most of the clump mass (282 M$_{\odot}$) is concentrated in the central region of the clump. Thus, these protostars are potential candidates to becoming high-mass stars by accreting gas material from the available mass reservoir. 

Using H$^{13}$CO$^{+}$ line emission, we derive the dispersion in the turbulent velocity along the line of sight, $\sigma v_\mathrm{los}(=\sqrt{\sigma_{\rm obs}^{2}-\sigma_{th}^{2}})$. The $\sigma_{\rm obs}$ and $\sigma_{th}$ are the total observed and thermal velocity dispersions, respectively. The $\sigma_{th}$ can be expressed as $\sqrt{k_{B}T/\mu m_{\rm H}}$, where $k_{B}$ is the Boltzmann constant, $\mu$ is the mean molecular weight per free particle (30). Assuming a temperature of 50 K, $\sigma_{th}$ and $\sigma v_\mathrm{los}$ are estimated as 0.12 and 1.21 km s$^{-1}$, respectively.

\subsection{Modeling of the hourglass-shaped magnetic field} \label{subsec:bpos_map}

In order to model the magnetic field pattern, we used the \textit{DustPol} module included in the ARTIST package \citep{2012A&A...543A..16P}, which is based on the Line Modeling Engine (LIME) radiative transfer code \citep{2010A&A...523A..25B}. \textit{DustPol} generates the synthetic Stokes $I$, $Q$, and $U$ images in FITS-format, which are used as an input for the \textit{simobserve} and \textit{simanalyze} tasks of CASA, considering the same antenna configuration of the observing runs in ALMA. In this way, polarization position angle maps are created and compared with the observations. The physical modeling of the 3D magnetic field is done by combining an axisymmetric singular toroid threaded by a poloidal field \citep{1996ApJ...472..211L,2011A&A...530A.109P}. Based on \cite{2013A&A...560A.114P}, we added a toroidal force-free component of the magnetic field to represent the effects of rotation. 

The model used in this work has four free parameters: (1) the mass-to-flux ratio normalized to its critical value, $\lambda$, defined as,
\begin{equation}
    \lambda=\frac{(M/\Phi_{B})}{(M/\Phi_{B})_{cr}}=2\pi\sqrt{G}\frac{M}{\Phi_{B}}
    \label{eq:lambda}
\end{equation}
\noindent
where $G$ is the gravitational constant, $\Phi_{B}$ is the magnetic flux ($=\pi R^{2}B$), and $M$ is the mass of the core; (2) the ratio of the strengths of the toroidal and the poloidal components of the magnetic field in the midplane of the source, $b_{0}$; (3) the orientation of the magnetic axis projected on the plane of sky, $\varphi$, starting from north and increasing eastward; (4) the inclination with respect to the plane of the sky, $i$, with an assumption to be positive/negative when the magnetic field in the northern part directs towards/away from us. A $\chi^{2}$-test is performed on the polarization angle residuals $(\Delta\psi)$ of the polarization angle $\psi$, obtained from the difference between the observed ($\psi_{\rm obs}$) and the modeled ($\psi_{\rm mod}$) polarization angles ($\Delta\psi=\psi_{\rm obs}-\psi_{\rm mod}$), enclosed within the circular area for each combination of the four free parameters. The $\chi^{2}$ is estimated accounting for regions lying above the 3$\sigma$ level ($\sigma=29 \mu$Jy beam$^{-1}$) of Stokes $Q$ and $U$ images. A constant temperature (50 K) has been adopted in the modelling as no temperature map of this region is available currently. On the other way, the region close to MM1 and MM2, is expected to be relatively hot and having more weight in estimating the Stokes parameters.

The best-fit model provides the minimum reduced $\chi^{2}$ value of 9.39 with the combination of model parameters: $\lambda=1.63$, $b_{0}=-0.1\pm0.1$, $i=45^{\circ+13^{\circ}}_{-6^{\circ}}$, and $\varphi=40^{\circ+1^{\circ}}_{-4^{\circ}}$. The errors are estimated following \cite{1976ApJ...208..177L}. A low value of $b_{0}$ in the best-fit model is consistent with no apparent rotation towards G333 as found in H$^{13}$CO$^{+}$. The value of $\lambda=1.63$ obtained from the model (see Equation \ref{eq:lambda}) is relative to the mass enclosed in a flux tube, but observationally the mass is derived considering a spherical system. Therefore, the estimated $\lambda$ is transformed to an effective $\lambda$ of 1.43 \citep{1996ApJ...472..211L}, which is marginally supercritical. 

\begin{figure}
\includegraphics[width=8.5cm, height=7.5cm]{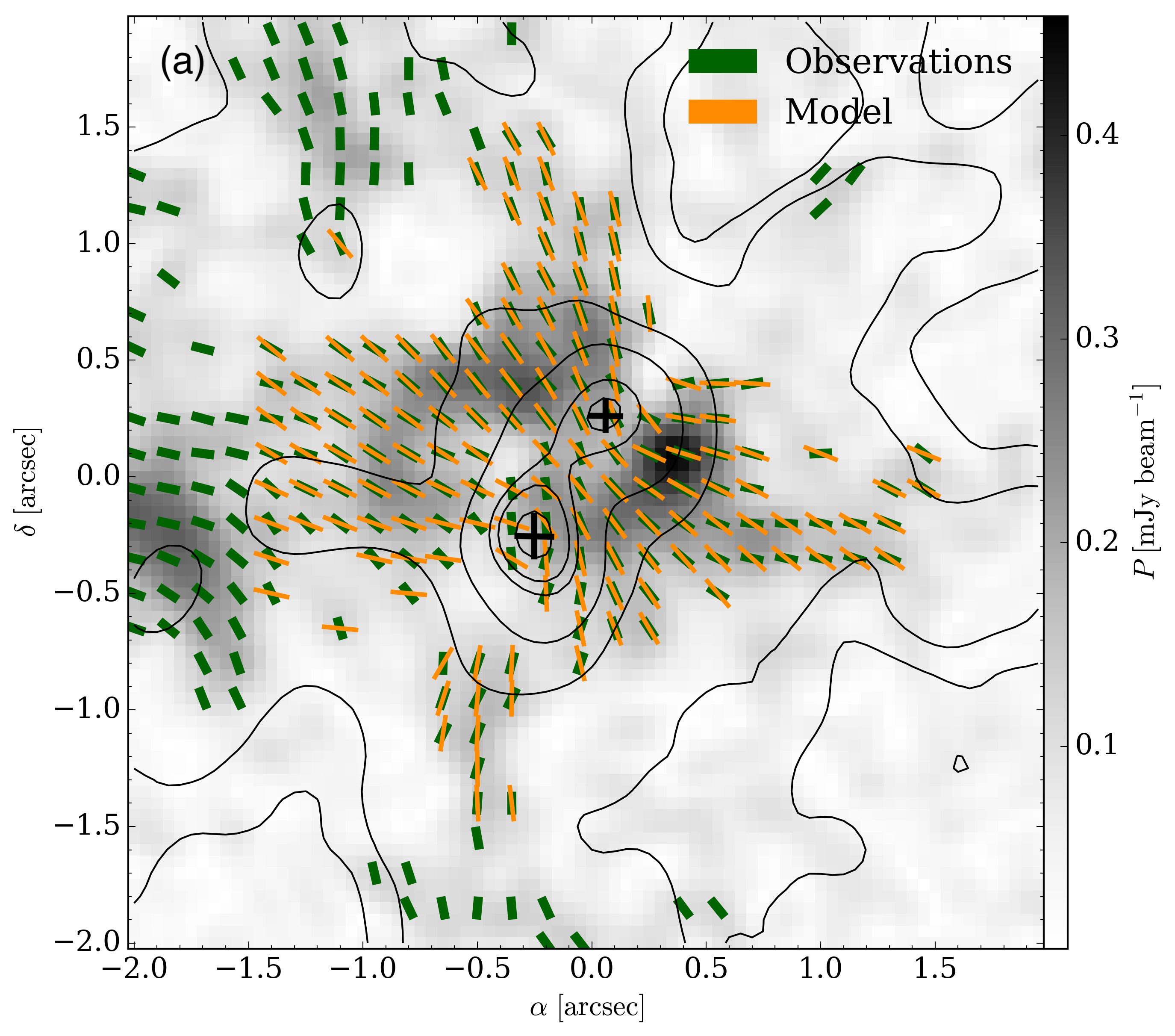}
\includegraphics[width=8.5cm, height=7.5cm]{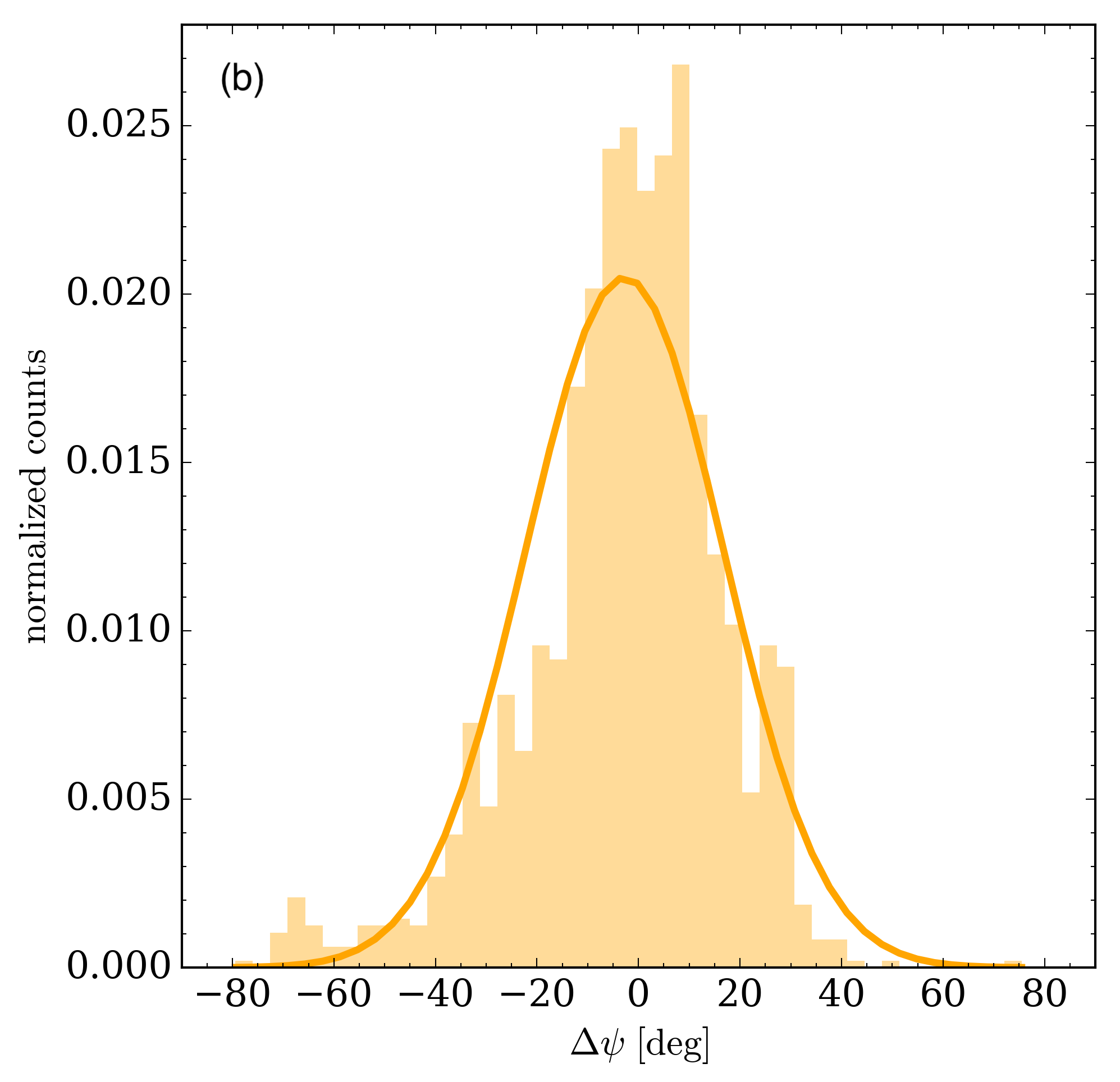}
\caption{\textbf{(a)} Comparison between the observed (green segments) and modeled (yellow segments) magnetic field orientations overplotted on the polarized intensity map (grayscale). The black contours are the same as Figure \ref{fig:bpos}. \textbf{(b)} Histogram of the polarization angle residuals ($\Delta\psi$) along with the Gaussian fitting.} 
\label{fig:bpos_fit}
\end{figure}

In Figure \ref{fig:bpos_fit} (a), we show a comparison between the observed and modeled magnetic field geometry, limited to the area within the circle shown in Figure \ref{fig:bpos} (a). Although the hourglass model reproduces quite well the observed magnetic field geometry, deviations can be seen around MM1, mainly towards its southern and eastern regions. Since MM1 is relatively more massive than MM2, it is possible that the gravity of MM1 might have dragged and distorted the magnetic field lines more effectively than MM2. In Figure \ref{fig:bpos_fit} (b), we show the histogram of the $\Delta\psi$. A Gaussian fit to the whole distribution of $\Delta\psi$ provides a mean value $\textless\Delta\psi\textgreater = -3.32^{\circ}$ and a standard deviation $\sigma_{\psi}=19.48^{\circ}$. 



\subsection{Magnetic field strength} \label{subsec:bpos_str}


We estimate B$_\mathrm{POS}$ towards G333 using the well-known Davis–Chandrasekhar–Fermi (DCF) method \citep{1951PhRv...81..890D,1953ApJ...118..113C}\footnote{also see \cite{2022ApJ...925...30L,2022FrASS...9.3556L}}, which states:
\begin{equation}
    B_\mathrm{POS} = Q\sqrt{4\pi\rho} \frac{\sigma v_\mathrm{los}}{\delta\psi_{int}}
\end{equation}
\noindent
where Q is a correction factor, 0.5,\footnote{In \cite{2001ApJ...546..980O} the size scale adopted is $\sim8$ pc, whereas \cite{2021ApJ...919...79L} estimated Q to be $\sim0.28$ at a size scale of a $1-0.2$ pc, which is relatively closer to the size scale of the area of analysis. Adopting the later, the \bpos~ is estimated as $3.0-3.6$ mG, which is consistent with our estimated \bpos.} adopted from simulations in turbulent clouds \citep{2001ApJ...546..980O}, $\rho$ is the mass density of the cloud, $\sigma v_\mathrm{los}$ is the dispersion in the turbulent velocity along the line of sight, and $\delta\psi_{int}$ is the intrinsic dispersion in the polarization angle $\psi$. The $\delta\psi_{int}$ can be obtained from $\sqrt{\sigma_{\psi}^{2}-\delta\psi_{\rm obs}^{2}}$. 
The average uncertainty in the observed $\psi$ within the circle of analysis is $\delta\psi_{\rm obs}=5.4^{\circ}$. The $\delta\psi_{int}$ obtained is $18.7^{\circ}$. 
Consequently, $B_\mathrm{pos}$ is estimated as 4.0 mG. Adopting $i=45^{\circ}$ from our model, we also estimate the total magnetic field strength $B=B_\mathrm{pos}/\mathrm{cos}~(i)=5.7$ mG. With a density of n(H$_{2})=8.4\times10^{6}$ cm$^{-3}$ this strength is consistent as found in the compilation of B-field strength with density by \cite{2022ApJ...925...30L} using the DCF method.

The DCF method assumes the deviation of uniform magnetic field by turbulent gas motions, without considering gravity, which is most likely responsible for the hourglass shape. In this work gravity is taken into account while using the \textit{dustPol} model and by considering the $\Delta\psi$, we eliminate the gravitational effect and the $\delta\psi$ is supposed to be the interplay between the turbulence and the magnetic field.

While computing $\lambda$ using Equation \ref{eq:lambda}, we assumed that all the mass is in the envelope, while the protostellar mass should also be added to the envelope mass. Therefore, the estimated $\lambda$ should be considered as a lower limit. As mentioned in \ref{subsec:mass}, based on the luminosity, an additional stellar mass of $\sim6$ M$_{\odot}$ could be added to the envelope mass, making the total mass of 29 M$_{\odot}$, which leads to a mass-to-flux ratio of 1.22. The fact that the dust emission may be optically thick at the very central bright fragmented region, leads to underestimate the mass of the circular region can not be ignored. However, the estimated value of $\lambda$ is similar to the one obtained from the modelling (1.43). 

We compute the Alfv$\mathrm{\acute{e}}$n speed $v_{\rm A}=B/\sqrt{4\pi\rho}=2.58$ km s$^{-1}$, through which we estimate the turbulent to magnetic energy $\beta_{t}=3(\sigma v_\mathrm{los}/v_{\rm A})^{2}=0.66$. This indicates that the magnetic energy dominates moderately over the turbulent energy in this region. Thus the magnetic field towards G333, in spite of being quite strong, could not prevent the fragmentation and eventually star formation. 

Based on a non-ideal magnetohydrodynamic simulations of an initially subcritical core by \cite{2020MNRAS.494..827M}, the collapse happens after the core reaches a marginally supercritical mass-to-flux ratio, leaving behind a considerable amount of magnetic flux, which helps to transport out the angular momentum through strong magnetic braking. As a result, the protostar may harbour no disc or a very small one, with a very weak outflow signature. In G333, we do not find any evidence of rotation in the envelope region (Figure \ref{fig:mom_maps} b), nor in any of the protostars at the current angular resolution. Even in higher angular resolution ($\sim0.05^{"}$) observations of the same region obtained as part of the Digging into the Interior of Hot Cores with the ALMA \citep[DIHCA;][]{2022ApJ...929...68O,2023ApJ...959L..31O,2023ApJ...950...57T} survey (P. Sanhueza, private communication), no clear signature of rotation was found in any of the protostars, making the argument of strong magnetic braking more evident. Additionally, we found no significant outflow signature coming from the two protostars in G333 (see appendix \ref{app:mol_outflow}). This is consistent with overall physical scenario of very weak outflow and no disk when starting with strongly magnetized initial conditions \citep{2020MNRAS.494..827M}. This leads to the possibility that star formation in G333 started in an initially subcritical or transcritical cloud. 



\subsection{Analysis of the Energy Balance}\label{subsec:energy_bal}

The current dynamical state towards G333 can be inferred from the virial parameter ($\alpha_\mathrm{vir}$), the ratio of the virial mass ($M_\mathrm{vir}$) to the total mass ($M$) of the system with considerations of both the turbulent and magnetic supports. While $\alpha_\mathrm{vir}\sim1$ indicates the equilibrium state, $\alpha_\mathrm{vir}\textgreater1$ implies expansion and conversely $\alpha_\mathrm{vir}\textless1$ signifies contraction under gravity. We estimate $\alpha_\mathrm{vir}$ as $1.3$ for a centrally peaked density profile (see Appendix \ref{app:virial_param}). This suggests that G333 is currently close to a quasi-equilibrium state or undergoing quasi-static evolution, which is further supported by the quiescent environment of G333 as implied by the velocity field traced by the H$^{13}$CO$^{+}$ line emission (Figure \ref{fig:mom_maps} b). A strongly magnetized environment exerts an adequate support against rapid gravitational collapse, although the magnetic field could not hinder the fragmentation. This is consistent with a second stage of fragmentation within an initial massive core that forms out of a transcritical cloud \citep{2014ApJ...780...40B}. 

\subsection{Flattened envelope harbouring the hourglass magnetic field}

Theoretical studies suggests that for a relatively strong magnetic field environment ($\lambda\leq3$), the core material is channeled primarily along the magnetic field lines by the gravitational pull, building a flattened pancake-like structure perpendicular to the field lines, and developing a small toroidal component of the magnetic field \citep{2003ApJ...599..363A,2007MNRAS.377...77P}. As shown in Figure \ref{fig:bpos}, the flattened structure where the fragmentation took place agrees well with the strong magnetic field case, which is further supported by a low $\lambda$ obtained from our best-fit model as well as the observations. Additionally a small contribution of the toroidal component of the field ($\sim10\%$ of the poloidal component) obtained from the best-fit model supports this phenomenon. This is consistent with some previous studies \citep[e.g.,][]{2014ApJ...794L..18Q,2019A&A...630A..54B, 2019ApJ...879...25K}, which suggest a major contribution of magnetic field in the formation of high-mass protostars.

\subsection{Stability analysis of the central protostars}

To investigate whether the central protostars MM1 and MM2 are gravitationally bound, we followed the methodology adopted in \cite{2015Natur.518..213P} and \cite{2024NatAs...8..472L}. We estimate the gravitational potential energy ($V_{i}$) by\\
\begin{equation}
    V_{i}=-\sum_{i\neq j} \frac{Gm_{i}m_{j}}{r_{ij}}
\end{equation} 
where $m_{i}$ and $m_{j}$ are the masses of objects $i$ and $j$, respectively, and $r_{ij}$ is the separation between them. The kinetic energy ($E_{i}$) is given by,
\begin{equation}
    E_{i}=\frac{1}{2}m_{i}(v_{i}-v_{com})^{2}
\end{equation}
where $v_{i}$ is the line-of-sight velocity of the object $i$, and $v_{com}$ is the velocity of the centre of mass of the system. We estimate $v_{com}$ by
\begin{equation}
    v_{com}=\frac{\sum_{k}{m_{k}v_{k}}}{\sum_{k}{m_{k}}}
\end{equation}

The velocities of MM1 and MM2 are obtained from \cite{2023ApJ...950...57T} as $-44.06$ and $-43.55$ km s$^{-1}$, respectively. Assuming that the measured velocity difference between MM1 and MM2 is in 1-D, the full velocity difference in 3-D is $\Delta v_{3D}=\sqrt{3}({v_{i}-v_{com}})$ along the line-of-sight. Also, the total separation between MM1 and MM2 is estimated by multiplying the measured projected separation (1740 au) with $4/\pi$, i.e., 2215 au, assuming a random orientation between them. 

We measured the masses of MM1 and MM2 from the flux density obtained from the \textit{astrodendro} task to estimate the $E_{i}/V_{i}$ for each of them. The $E_{i}/V_{i}$ for MM1 and MM2 are estimated as 0.08 and 0.17, respectively. As both MM1 and MM2 show $E_{i}/V_{i}\textless1$, they could be gravitationally bound in a binary system at the present stage.
\\\\


\section{Summary and Conclusions} \label{sec:con}

The ALMA high-angular resolution ($\sim0.3^{\prime\prime}$) observations of linearly polarized 1.2 mm dust emission towards the high-mass star-forming region G333.46$-$0.16 revealed an hourglass-shaped pattern of B$_\mathrm{pos}$ at a scale of $\sim6000$ au. 
The hourglass shape is found to be more pinched towards MM1 than MM2, which might be because of the larger mass of MM1 compared to MM2. The protostars MM1 and MM2 are found to be gravitationally bound in a binary system at present with a separation of $1740$ au.

The H$^{13}$CO$^{+}$ line emission shows no strong velocity gradient that could hint any rotation towards G333. 
Also, none of the protostars show any clear signature of outflows, suggesting a strong magnetic braking. 

The hourglass-shaped \bpos~ is well fitted with an axisymmetric semi-analytical magnetostatic model. The best fit model is primarily governed by a poloidal component tangled with a small toroidal component ($\sim10\%$ of the poloidal one). The best fit results also provide $\lambda=1.43$, $i=45^{\circ+13^{\circ}}_{-6^{\circ}}$ and $\phi=40^{\circ+1^{\circ}}_{-4^{\circ}}$. 

Based on the DCF relation, the total magnetic field strength is estimated as $5.7$ mG. We also estimate the mass-to-flux ratio using the dispersion of polarization angle obtained from the model fitting, which results in $1.22$, consistent with same obtained from the best fit model ($\lambda=1.43$). 

Our analysis of energy balance in this area suggests a quasi-equilibrium state, which is consistent with the current dynamic environment inferred by H$^{13}$CO$^{+}$ line emission. Our speculation based on this work is that the magnetic field towards G333 is strong enough to stabilize the environment towards the center. A low magnitude of turbulent-to-magnetic energy ratio indicates a suppression of the magnetic field over the turbulence in this region. However, both are ultimately overwhelmed by the gravity, responsible for the fragmentation and eventually star formation. This leads to a possible scenario that the star formation in G333 might have started in an initially subcritical core, when the very central peak density area reached the marginally supercritical mass-to-flux ratio.


\begin{acknowledgments}
We thank the anonymous referee for providing constructive suggestions which enriched the draft significantly. P.S. was partially supported by a Grant-in-Aid for Scientific Research (KAKENHI numbers JP22H01271 and JP24K17100) of the Japan Society for the Promotion of Science (JSPS). P.S. was partially supported by a Grant-in-Aid for Scientific Research (KAKENHI Numbers JP22H01271 and JP23H01221) of the JSPS. J.M.G., acknowledges support by grant PID2020-117710GB-I00 (MCI-AEI-FEDER, UE). This work is also partially supported by the program Unidad de Excelencia María de Maeztu CEX2020-001058-M. J.L. is partially supported by a Grant-in-Aid for Scientific Research (KAKENHI Number JP23H01221) of JSPS. A.S.-M.\ acknowledges support from the RyC2021-032892-I grant funded by MCIN/AEI/10.13039/501100011033 and by the European Union `Next GenerationEU'/PRTR, as well as the program Unidad de Excelencia María de Maeztu CEX2020-001058-M, and support from the PID2020-117710GB-I00 (MCI-AEI-FEDER, UE). M.T.B. acknowledges financial support through the INAF Large Grant {\it The role of MAGnetic fields in MAssive star formation} (MAGMA). This paper makes use of the following ALMA data: ADS/JAO.ALMA\#2017.1.00101.S. K.P. is a Royal Society University Research Fellow, supported by grant number URF$\setminus$R1$\setminus$211322. XL acknowledges support from the National Key R\&D Program of China (No.\ 2022YFA1603101), the Natural Science Foundation of Shanghai (No.\ 23ZR1482100), the National Natural Science Foundation of China (NSFC) through grant Nos. 12273090 \& 12322305, and the Chinese Academy of Sciences (CAS) `Light of West China' Program (No.\ xbzgzdsys-202212). T.Cs. has received financial support from the French State in the framework of the IdEx Universit\'e de Bordeaux Investments for the future Program. Y.C. was partially supported by a Grant-in-Aid for Scientific Research (KAKENHI number JP24K17103) of the JSPS. This paper makes use of the following ALMA data: ADS/JAO. Data analysis was [in part] carried out on the Multi-wavelength Data Analysis System operated by the Astronomy Data Center (ADC), National Astronomical Observatory of Japan. ALMA is a partnership of ESO (representing its member states), NSF (USA) and NINS (Japan), together with NRC (Canada), MOST and ASIAA (Taiwan), and KASI (Republic of Korea), in cooperation with the Republic of Chile. The Joint ALMA Observatory is operated by ESO, AUI/NRAO and NAOJ. This research made use of astrodendro, a Python package to compute dendrograms of Astronomical data (\url{http://www.dendrograms.org/}).
\end{acknowledgments}

%
\facility{ALMA}
\software{CASA \citep{2007ASPC..376..127M,2022PASP..134k4501C}}

\appendix

\begin{figure}
\includegraphics[width=8.5cm, height=5.7cm]{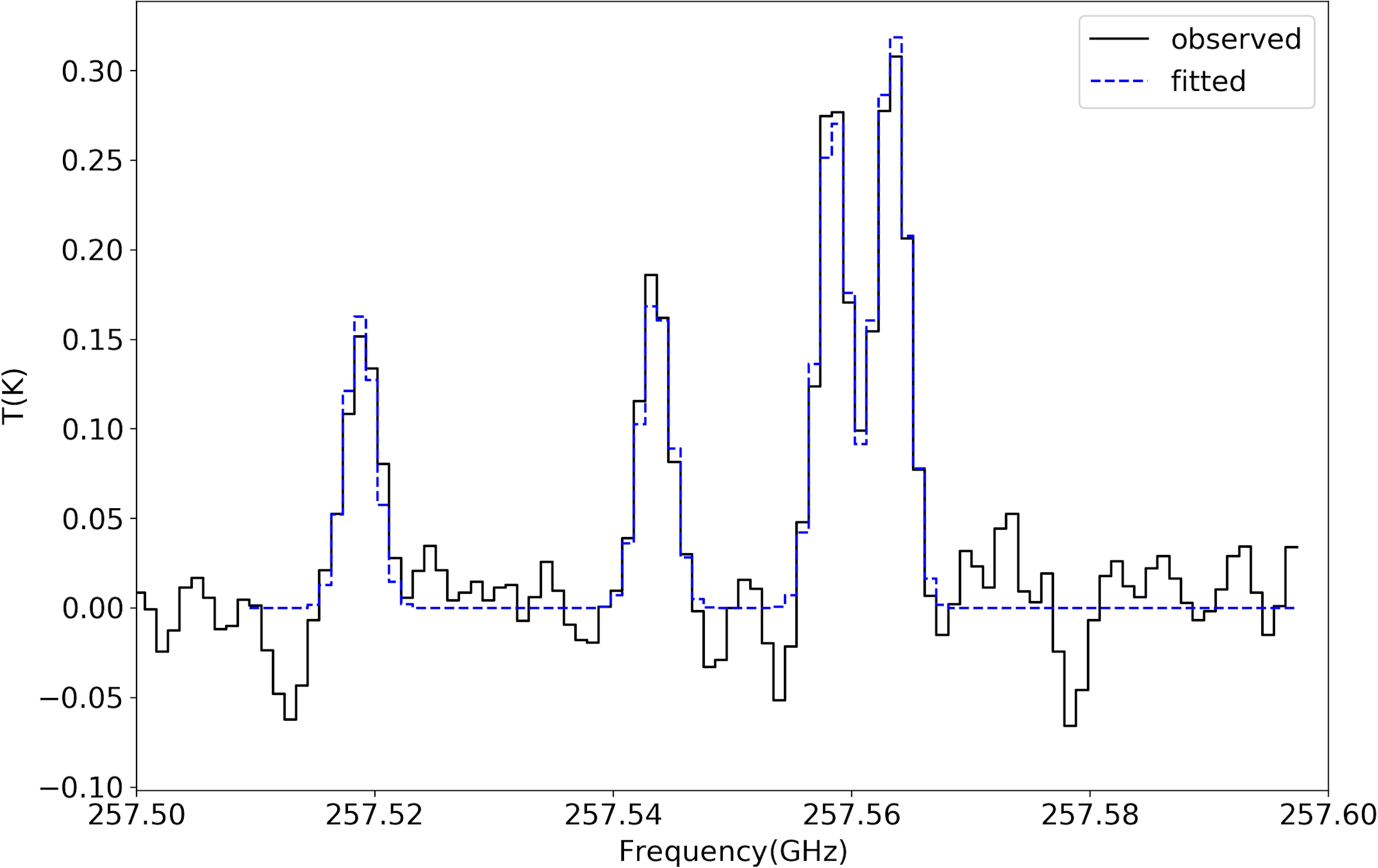}
\caption{The spectra of CH$_{3}$CN ($J=14-13$) towards G333. The black line shows the observed spectrum and the dashed blue line shows the fitted spectrum.}
\label{fig:ch3cn_temp}
\end{figure}

\begin{table}[]
    \scriptsize
    \begin{center}
    \caption{Results obtained from \textit{astrodendro}$^{*}$}
    \begin{tabular}{ccccc}
    \hline
     R.A.  &   Dec. &   FWHM    & Peak Intensity    &   Flux Density\\
        (ICRS)  &   (ICRS)  &   $(^{\prime\prime}\times^{\prime\prime})$    &   (mJy beam$^{-1}$)    &   (mJy)\\
        \hline
        \multicolumn{5}{c}{MM1}\\
         16:21:20.20    &   -50:09:46.91    &   $0.31\times0.21$  &   53.5  &   59.5\\
         \multicolumn{5}{c}{MM2}\\
         16:21:20.17    &   -50:09:46.41    &   $0.23\times0.14$    &  38.8  &   25.8\\
        \hline
    \end{tabular}
    \label{tab:dedro}
    \end{center}
    $^{*}$ $F_{min}=5\sigma$: the threshold above which the structures are defined.\\
    $\delta=1\sigma$: the minimum significance for separation of the structures. \\
    $S_{min}=$half of the beamsize: the minimum area to be allocated in the individual structure.
\end{table}

\section{Molecular outflows identified around G333}\label{app:mol_outflow}

\begin{figure}
\includegraphics[width=8.5cm, height=7cm]{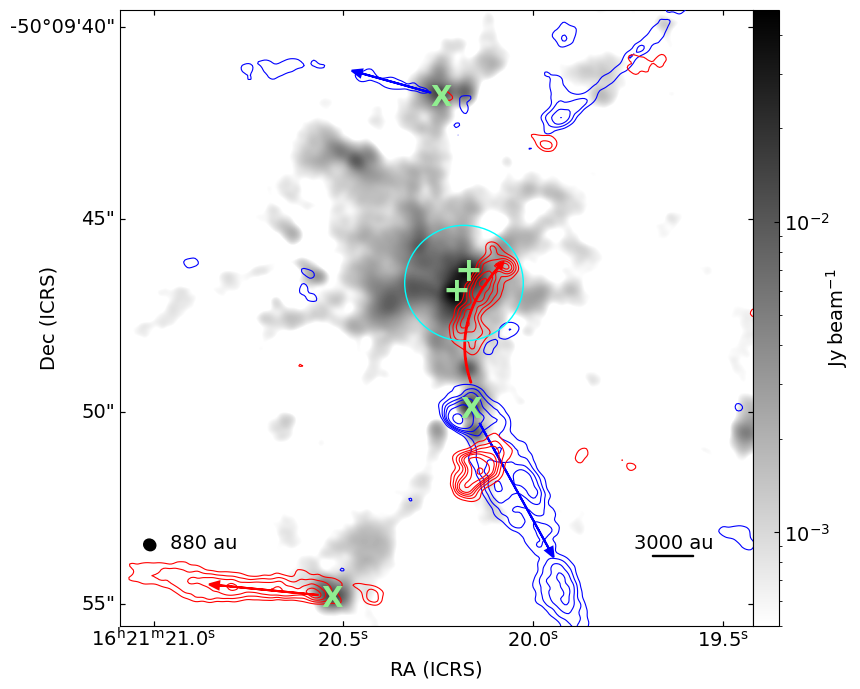}
\caption{Molecular outflows identified in $^{12}$CO $(J=3-2)$ overplotted on our observed 1.2 mm dust continuum emission. The plus symbols mark the positions of MM1 and MM2. The circle in cyan shows the region of analysis. The blueshifted and redshifted outflows are shown in blue and red contours, respectively. The blue contours are drawn with levels 3,5,7,9,12, and 18 times 0.13 Jy beam$^{-1}$. The red contours are drawn with levels 3,5,7,9,11,15,20, and 25 times 0.10 Jy beam$^{-1}$. The possible sources driving the ouflows are marked using green `X' symbols. The red and blue arrows show the directions of the redshifted and blueshifted outflows, respectively.}
\label{fig:outflow}
\end{figure}

Several molecular outflows are identified around G333 using the $^{12}$CO ($J=3-2$) line emission observed by ALMA 
(Project ID: 2013.1.00960.S, PI: T. Csengeri). 
Using the CASA task \textit{imregrid}, we match the phase-center of the $^{12}$CO line image and our 1.2 mm dust continuum image.

Figure \ref{fig:outflow} shows the $^{12}$CO molecular outflows overplotted on the dust continuum image. The $^{12}$CO line emission is integrated from $-106.1$ to $-65.3$ km s$^{-1}$ for the blueshifted lobe, and $-22.1$ to $0.7$ km s$^{-1}$ for the redshifted lobe. The sensitivities of the integrated images of the blueshifted and redshifted $^{12}$CO line emission are 0.14 and 0.10 Jy beam$^{-1}$, respectively. The outflow lobes in the north and south of the region enclosed by the hourglass-shaped magnetic field (shown as a cyan colored circle in Figure \ref{fig:outflow}) are highly collimated. Interestingly, it is noticeable that although there is a significant amount of redshifted $^{12}$CO line emission within the circle (which seems to be the redshifted lobe of the nearby dense core located south to the circular area of analysis), none of the outflow lobes is driven by any of the protostars but likely the low-mass protostars residing outskirts of the dense gas. There are some dust condensations that likely could host these protostars, marked with `X' symbols in Figure \ref{fig:outflow}. 

The absence of outflows in any of the protostars makes it difficult to predict the of angular momentum direction. 
In a classical scenario of magnetized star formation, the angular momentum is efficiently lost by magnetic braking \citep{1979ApJ...230..204M}. Our speculation is that magnetic braking is relatively strong in the case of G333, 
because of which we do not find any observational signature of 
strong angular momentum.

\section{Virial parameter}\label{app:virial_param}

The virial theorem can be expressed as below:
\begin{equation}
    \frac{1}{2}\frac{d^{2}I}{dt^{2}}=2E_{k}+E_{G}+E_{B}
\end{equation}
\noindent
where $I$ is the moment of inertia, $E_{k}$, $E_{G}$, and $E_{B}$ are the kinetic, gravitational, and magnetic energy, respectively. We consider a spherical cloud having a radial density profile $\rho\propto r^{-\alpha}$, where $\alpha=2$ imply the density profile to be centrally peaked \citep{2013EAS....62...25B}. The gravitational energy ($E_{G}$) can be expressed by:
\begin{equation}
    E_{G} = -\frac{(3-\alpha)}{(5-2\alpha)}\frac{GM^{2}}{R}
\end{equation}
\noindent
The magnetic energy ($E_{B}$) is given by:
\begin{equation}
    E_{B} = \frac{B^{2}V}{8\pi}
\end{equation}
\noindent
The kinetic energy ($E_{k}$) is derived by:
\begin{equation}
    E_{k} = -\frac{3}{2}M\sigma_{\rm obs}^{2}
\end{equation}
The rotational energy ($E_{r}$) is expressed as:
\begin{equation}
    E_{r} = \frac{1}{3}Mv_{\rm rot}^{2}\frac{(3-\alpha)}{(5-\alpha)}
\end{equation}
We do not find any evidence of rotation with a spectral resolution 0.56 km s$^{-1}$ in the area of analysis. If we assume this resolution as the upper limit of the velocity due to rotation ($v_{\rm rot}$), we estimate the ratio $E_{rot}/E_{G}$ to be 0.006 for a centrally peak density profile. With the same assumption of rotation, a ratio of $E_{rot}/E_{B}$ is estimated to be 0.013 for the same profile. Similarly, we estimate the ratio of $E_{rot}/E_{k}$ as 0.014. As the contribution of the rotational energy is negligible compared to the other energies, we do not account the upper limit of rotational term in the estimation of virial parameter.

Including both the kinetic and magnetic energies, a system can be in stable phase, when $2E_{k}+E_{B}+E_{G}\textless0$. The ratio between the virial mass $M_\mathrm{vir} (=M_{k+B})$ and the total mass $M$ can be defined as the virial parameter $\alpha_\mathrm{vir}$. Following \cite{2020ApJ...895..142L}, $\alpha_\mathrm{vir}$ is given below:
\begin{equation}
    \alpha_\mathrm{vir} = \frac{M_\mathrm{k+B}}{M}=\frac{1}{M}\Biggr[\sqrt{M_\mathrm{B}^{2}+\Biggl(\frac{M_\mathrm{k}^{2}}{2}}\Biggr)+\frac{M_\mathrm{k}}{2}\Biggl]
\end{equation}
where $M_\mathrm{k+B}$ is the total virial mass considering the kinetic motion and magnetic field. The kinetic virial mass $M_\mathrm{k}$ is defined as,
\begin{equation}
    M_\mathrm{k} = 3\Biggl(\frac{5-2\alpha}{3-\alpha}\Biggr)\frac{\sigma_\mathrm{los}^{2}R}{G}
\end{equation}
\\\\
The virial mass accounting for the ordered magnetic field $M_\mathrm{B}$ is expressed as:
\begin{equation}
    M_\mathrm{B} = \frac{\pi R^{2}B}{\sqrt{\frac{6(3-\alpha)}{(5-2\alpha)}\pi^{2} G}}    
\end{equation}
\noindent

We estimate the $\alpha_\mathrm{vir}$ to be $1.3$, which suggests an equilibrium state of G333, implying that magnetic field provides a significant support against gravity.


\bibliography{sample631}{}

\begin{thebibliography}{}
\expandafter\ifx\csname natexlab\endcsname\relax\def\natexlab#1{#1}\fi
\providecommand{\url}[1]{\href{#1}{#1}}
\providecommand{\dodoi}[1]{doi:~\href{http://doi.org/#1}{\nolinkurl{#1}}}
\providecommand{\doeprint}[1]{\href{http://ascl.net/#1}{\nolinkurl{http://ascl.net/#1}}}
\providecommand{\doarXiv}[1]{\href{https://arxiv.org/abs/#1}{\nolinkurl{https://arxiv.org/abs/#1}}}

\bibitem[{{Allen} {et~al.}(2003){Allen}, {Li}, \& {Shu}}]{2003ApJ...599..363A}
{Allen}, A., {Li}, Z.-Y., \& {Shu}, F.~H. 2003, \apj, 599, 363, \dodoi{10.1086/379243}

\bibitem[{{Andersson} {et~al.}(2015){Andersson}, {Lazarian}, \& {Vaillancourt}}]{2015ARA&A..53..501A}
{Andersson}, B.~G., {Lazarian}, A., \& {Vaillancourt}, J.~E. 2015, \araa, 53, 501, \dodoi{10.1146/annurev-astro-082214-122414}

\bibitem[{{Bailey} \& {Basu}(2014)}]{2014ApJ...780...40B}
{Bailey}, N.~D., \& {Basu}, S. 2014, \apj, 780, 40, \dodoi{10.1088/0004-637X/780/1/40}

\bibitem[{{Belloche}(2013)}]{2013EAS....62...25B}
{Belloche}, A. 2013, in EAS Publications Series, Vol.~62, EAS Publications Series, ed. P.~{Hennebelle} \& C.~{Charbonnel}, 25--66, \dodoi{10.1051/eas/1362002}

\bibitem[{{Beltr{\'a}n} {et~al.}(2019){Beltr{\'a}n}, {Padovani}, {Girart}, {Galli}, {Cesaroni}, {Paladino}, {Anglada}, {Estalella}, {Osorio}, {Rao}, {S{\'a}nchez-Monge}, \& {Zhang}}]{2019A&A...630A..54B}
{Beltr{\'a}n}, M.~T., {Padovani}, M., {Girart}, J.~M., {et~al.} 2019, \aap, 630, A54, \dodoi{10.1051/0004-6361/201935701}

\bibitem[{{Brinch} \& {Hogerheijde}(2010)}]{2010A&A...523A..25B}
{Brinch}, C., \& {Hogerheijde}, M.~R. 2010, \aap, 523, A25, \dodoi{10.1051/0004-6361/201015333}

\bibitem[{{CASA Team} {et~al.}(2022){CASA Team}, {Bean}, {Bhatnagar}, {Castro}, {Donovan Meyer}, {Emonts}, {Garcia}, {Garwood}, {Golap}, {Gonzalez Villalba}, {Harris}, {Hayashi}, {Hoskins}, {Hsieh}, {Jagannathan}, {Kawasaki}, {Keimpema}, {Kettenis}, {Lopez}, {Marvil}, {Masters}, {McNichols}, {Mehringer}, {Miel}, {Moellenbrock}, {Montesino}, {Nakazato}, {Ott}, {Petry}, {Pokorny}, {Raba}, {Rau}, {Schiebel}, {Schweighart}, {Sekhar}, {Shimada}, {Small}, {Steeb}, {Sugimoto}, {Suoranta}, {Tsutsumi}, {van Bemmel}, {Verkouter}, {Wells}, {Xiong}, {Szomoru}, {Griffith}, {Glendenning}, \& {Kern}}]{2022PASP..134k4501C}
{CASA Team}, {Bean}, B., {Bhatnagar}, S., {et~al.} 2022, \pasp, 134, 114501, \dodoi{10.1088/1538-3873/ac9642}

\bibitem[{{Chandrasekhar} \& {Fermi}(1953)}]{1953ApJ...118..113C}
{Chandrasekhar}, S., \& {Fermi}, E. 1953, \apj, 118, 113, \dodoi{10.1086/145731}

\bibitem[{{Contreras} {et~al.}(2018){Contreras}, {Sanhueza}, {Jackson}, {Guzm{\'a}n}, {Longmore}, {Garay}, {Zhang}, {Nguyễn-Lu'o'ng}, {Tatematsu}, {Nakamura}, {Sakai}, {Ohashi}, {Liu}, {Saito}, {Gomez}, {Rathborne}, \& {Whitaker}}]{2018ApJ...861...14C}
{Contreras}, Y., {Sanhueza}, P., {Jackson}, J.~M., {et~al.} 2018, \apj, 861, 14, \dodoi{10.3847/1538-4357/aac2ec}

\bibitem[{{Cort{\'e}s} {et~al.}(2021){Cort{\'e}s}, {Sanhueza}, {Houde}, {Mart{\'\i}n}, {Hull}, {Girart}, {Zhang}, {Fernandez-Lopez}, {Zapata}, {Stephens}, {Li}, {Wu}, {Olguin}, {Lu}, {Guzm{\'a}n}, \& {Nakamura}}]{2021ApJ...923..204C}
{Cort{\'e}s}, P.~C., {Sanhueza}, P., {Houde}, M., {et~al.} 2021, \apj, 923, 204, \dodoi{10.3847/1538-4357/ac28a1}

\bibitem[{{Csengeri} {et~al.}(2014){Csengeri}, {Urquhart}, {Schuller}, {Motte}, {Bontemps}, {Wyrowski}, {Menten}, {Bronfman}, {Beuther}, {Henning}, {Testi}, {Zavagno}, \& {Walmsley}}]{2014A&A...565A..75C}
{Csengeri}, T., {Urquhart}, J.~S., {Schuller}, F., {et~al.} 2014, \aap, 565, A75, \dodoi{10.1051/0004-6361/201322434}

\bibitem[{{Csengeri} {et~al.}(2017){Csengeri}, {Bontemps}, {Wyrowski}, {Motte}, {Menten}, {Beuther}, {Bronfman}, {Commer{\c{c}}on}, {Chapillon}, {Duarte-Cabral}, {Fuller}, {Henning}, {Leurini}, {Longmore}, {Palau}, {Peretto}, {Schuller}, {Tan}, {Testi}, {Traficante}, \& {Urquhart}}]{2017A&A...600L..10C}
{Csengeri}, T., {Bontemps}, S., {Wyrowski}, F., {et~al.} 2017, \aap, 600, L10, \dodoi{10.1051/0004-6361/201629754}

\bibitem[{{Cudlip} {et~al.}(1982){Cudlip}, {Furniss}, {King}, \& {Jennings}}]{1982MNRAS.200.1169C}
{Cudlip}, W., {Furniss}, I., {King}, K.~J., \& {Jennings}, R.~E. 1982, \mnras, 200, 1169, \dodoi{10.1093/mnras/200.4.1169}

\bibitem[{{Davis}(1951)}]{1951PhRv...81..890D}
{Davis}, L. 1951, Physical Review, 81, 890, \dodoi{10.1103/PhysRev.81.890.2}

\bibitem[{{Estalella} {et~al.}(2019){Estalella}, {Anglada}, {D{\'\i}az-Rodr{\'\i}guez}, \& {Mayen-Gijon}}]{2019A&A...626A..84E}
{Estalella}, R., {Anglada}, G., {D{\'\i}az-Rodr{\'\i}guez}, A.~K., \& {Mayen-Gijon}, J.~M. 2019, \aap, 626, A84, \dodoi{10.1051/0004-6361/201834998}

\bibitem[{{Fern{\'a}ndez-L{\'o}pez} {et~al.}(2021){Fern{\'a}ndez-L{\'o}pez}, {Sanhueza}, {Zapata}, {Stephens}, {Hull}, {Zhang}, {Girart}, {Koch}, {Cort{\'e}s}, {Silva}, {Tatematsu}, {Nakamura}, {Guzm{\'a}n}, {Nguyen Luong}, {Guzm{\'a}n Ccolque}, {Tang}, \& {Chen}}]{2021ApJ...913...29F}
{Fern{\'a}ndez-L{\'o}pez}, M., {Sanhueza}, P., {Zapata}, L.~A., {et~al.} 2021, \apj, 913, 29, \dodoi{10.3847/1538-4357/abf2b6}

\bibitem[{{Girart} {et~al.}(2009){Girart}, {Beltr{\'a}n}, {Zhang}, {Rao}, \& {Estalella}}]{2009Sci...324.1408G}
{Girart}, J.~M., {Beltr{\'a}n}, M.~T., {Zhang}, Q., {Rao}, R., \& {Estalella}, R. 2009, Science, 324, 1408, \dodoi{10.1126/science.1171807}

\bibitem[{{Girart} {et~al.}(2006){Girart}, {Rao}, \& {Marrone}}]{2006Sci...313..812G}
{Girart}, J.~M., {Rao}, R., \& {Marrone}, D.~P. 2006, Science, 313, 812, \dodoi{10.1126/science.1129093}

\bibitem[{{Hennebelle} \& {Inutsuka}(2019)}]{2019FrASS...6....5H}
{Hennebelle}, P., \& {Inutsuka}, S.-i. 2019, Frontiers in Astronomy and Space Sciences, 6, 5, \dodoi{10.3389/fspas.2019.00005}

\bibitem[{{Hildebrand}(1988)}]{1988QJRAS..29..327H}
{Hildebrand}, R.~H. 1988, \qjras, 29, 327

\bibitem[{{Hildebrand} {et~al.}(1984){Hildebrand}, {Dragovan}, \& {Novak}}]{1984ApJ...284L..51H}
{Hildebrand}, R.~H., {Dragovan}, M., \& {Novak}, G. 1984, \apjl, 284, L51, \dodoi{10.1086/184351}

\bibitem[{{Huang} {et~al.}(2024){Huang}, {Girart}, {Stephens}, {Fern{\'a}ndez L{\'o}pez}, {Arce}, {Carpenter}, {Cortes}, {Cox}, {Friesen}, {Le Gouellec}, {Hull}, {Karnath}, {Kwon}, {Li}, {Looney}, {Megeath}, {Myers}, {Murillo}, {Pineda}, {Sadavoy}, {S{\'a}nchez-Monge}, {Sanhueza}, {Tobin}, {Zhang}, {Jackson}, \& {Segura-Cox}}]{2024ApJ...963L..31H}
{Huang}, B., {Girart}, J.~M., {Stephens}, I.~W., {et~al.} 2024, \apjl, 963, L31, \dodoi{10.3847/2041-8213/ad27d4}

\bibitem[{{Hull} {et~al.}(2014){Hull}, {Plambeck}, {Kwon}, {Bower}, {Carpenter}, {Crutcher}, {Fiege}, {Franzmann}, {Hakobian}, {Heiles}, {Houde}, {Hughes}, {Lamb}, {Looney}, {Marrone}, {Matthews}, {Pillai}, {Pound}, {Rahman}, {Sandell}, {Stephens}, {Tobin}, {Vaillancourt}, {Volgenau}, \& {Wright}}]{2014ApJS..213...13H}
{Hull}, C. L.~H., {Plambeck}, R.~L., {Kwon}, W., {et~al.} 2014, \apjs, 213, 13, \dodoi{10.1088/0067-0049/213/1/13}

\bibitem[{{Hwang} {et~al.}(2022){Hwang}, {Kim}, {Pattle}, {Lee}, {Koch}, {Johnstone}, {Tomisaka}, {Whitworth}, {Furuya}, {Kang}, {Lyo}, {Chung}, {Arzoumanian}, {Park}, {Kwon}, {Kim}, {Tamura}, {Kwon}, {Soam}, {Han}, {Hoang}, {Kim}, {Onaka}, {Eswaraiah}, {Ward-Thompson}, {Liu}, {Tang}, {Chen}, {Matsumura}, {Hoang}, {Chen}, {Le Gouellec}, {Kirchschlager}, {Poidevin}, {Bastien}, {Qiu}, {Hasegawa}, {Lai}, {Byun}, {Cho}, {Choi}, {Choi}, {Choi}, {Jeong}, {Kang}, {Kim}, {Kim}, {Lee}, {Lee}, {Lee}, {Lee}, {Kim}, {Yoo}, {Yun}, {Chen}, {Di Francesco}, {Fiege}, {Fissel}, {Franzmann}, {Houde}, {Lacaille}, {Matthews}, {Sadavoy}, {Moriarty-Schieven}, {Tahani}, {Ching}, {Dai}, {Duan}, {Gu}, {Law}, {Li}, {Li}, {Li}, {Li}, {Liu}, {Lu}, {Qian}, {Wang}, {Wu}, {Xie}, {Yuan}, {Zhang}, {Zhang}, {Zhang}, {Zhou}, {Zhu}, {Berry}, {Friberg}, {Graves}, {Liu}, {Mairs}, {Parsons}, {Rawlings}, {Doi}, {Hayashi}, {Hull}, {Inoue}, {Inutsuka}, {Iwasaki}, {Kataoka}, {Kawabata}, {Kim}, {Kobayashi}, {Nagata}, {Nakamura}, {Nakanishi}, {Pyo},
  {Saito}, {Seta}, {Shimajiri}, {Shinnaga}, {Tsukamoto}, {Zenko}, {Chen}, {Duan}, {Fanciullo}, {Kemper}, {Lee}, {Lin}, {Liu}, {Ohashi}, {Rao}, {Tang}, {Wang}, {Yang}, {Yen}, {Bourke}, {Chrysostomou}, {Debattista}, {Eden}, {Eyres}, {Falle}, {Fuller}, {Gledhill}, {Greaves}, {Griffin}, {Hatchell}, {Karoly}, {Kirk}, {K{\"o}nyves}, {Longmore}, {van Loo}, {de Looze}, {Peretto}, {Priestley}, {Rawlings}, {Retter}, {Richer}, {Rigby}, {Savini}, {Scaife}, {Viti}, {Diep}, {Ngoc}, {Tram}, {Andr{\'e}}, {Coud{\'e}}, {Dowell}, {Friesen}, \& {Robitaille}}]{2022ApJ...941...51H}
{Hwang}, J., {Kim}, J., {Pattle}, K., {et~al.} 2022, \apj, 941, 51, \dodoi{10.3847/1538-4357/ac99e0}

\bibitem[{{Kirk} {et~al.}(2013){Kirk}, {Ward-Thompson}, {Palmeirim}, {Andr{\'e}}, {Griffin}, {Hargrave}, {K{\"o}nyves}, {Bernard}, {Nutter}, {Sibthorpe}, {Di Francesco}, {Abergel}, {Arzoumanian}, {Benedettini}, {Bontemps}, {Elia}, {Hennemann}, {Hill}, {Men'shchikov}, {Motte}, {Nguyen-Luong}, {Peretto}, {Pezzuto}, {Rygl}, {Sadavoy}, {Schisano}, {Schneider}, {Testi}, \& {White}}]{2013MNRAS.432.1424K}
{Kirk}, J.~M., {Ward-Thompson}, D., {Palmeirim}, P., {et~al.} 2013, \mnras, 432, 1424, \dodoi{10.1093/mnras/stt561}

\bibitem[{{Koch} {et~al.}(2018){Koch}, {Tang}, {Ho}, {Yen}, {Su}, \& {Takakuwa}}]{2018ApJ...855...39K}
{Koch}, P.~M., {Tang}, Y.-W., {Ho}, P. T.~P., {et~al.} 2018, \apj, 855, 39, \dodoi{10.3847/1538-4357/aaa4c1}

\bibitem[{{Kwon} {et~al.}(2019){Kwon}, {Stephens}, {Tobin}, {Looney}, {Li}, {van der Tak}, \& {Crutcher}}]{2019ApJ...879...25K}
{Kwon}, W., {Stephens}, I.~W., {Tobin}, J.~J., {et~al.} 2019, \apj, 879, 25, \dodoi{10.3847/1538-4357/ab24c8}

\bibitem[{{Lampton} {et~al.}(1976){Lampton}, {Margon}, \& {Bowyer}}]{1976ApJ...208..177L}
{Lampton}, M., {Margon}, B., \& {Bowyer}, S. 1976, \apj, 208, 177, \dodoi{10.1086/154592}

\bibitem[{{Lazarian}(2000)}]{2000ASPC..215...69L}
{Lazarian}, A. 2000, in Astronomical Society of the Pacific Conference Series, Vol. 215, Cosmic Evolution and Galaxy Formation: Structure, Interactions, and Feedback, ed. J.~{Franco}, L.~{Terlevich}, O.~{L{\'o}pez-Cruz}, \& I.~{Aretxaga}, 69, \dodoi{10.48550/arXiv.astro-ph/0003314}

\bibitem[{{Li} {et~al.}(2017){Li}, {Jiang}, {Fan}, {Gu}, \& {Zhang}}]{2017NatAs...1E.158L}
{Li}, H.-B., {Jiang}, H., {Fan}, X., {Gu}, Q., \& {Zhang}, Y. 2017, Nature Astronomy, 1, 0158, \dodoi{10.1038/s41550-017-0158}

\bibitem[{{Li} {et~al.}(2015){Li}, {Yuen}, {Otto}, {Leung}, {Sridharan}, {Zhang}, {Liu}, {Tang}, \& {Qiu}}]{2015Natur.520..518L}
{Li}, H.-B., {Yuen}, K.~H., {Otto}, F., {et~al.} 2015, \nat, 520, 518, \dodoi{10.1038/nature14291}

\bibitem[{{Li} {et~al.}(2024){Li}, {Sanhueza}, {Beuther}, {Chen}, {Kuiper}, {Olguin}, {Pudritz}, {Stephens}, {Zhang}, {Nakamura}, {Lu}, {Kuruwita}, {Sakai}, {Henning}, {Taniguchi}, \& {Li}}]{2024NatAs...8..472L}
{Li}, S., {Sanhueza}, P., {Beuther}, H., {et~al.} 2024, Nature Astronomy, 8, 472, \dodoi{10.1038/s41550-023-02181-9}

\bibitem[{{Li} \& {Shu}(1996)}]{1996ApJ...472..211L}
{Li}, Z.-Y., \& {Shu}, F.~H. 1996, \apj, 472, 211, \dodoi{10.1086/178056}

\bibitem[{{Lin} {et~al.}(2019){Lin}, {Csengeri}, {Wyrowski}, {Urquhart}, {Schuller}, {Weiss}, \& {Menten}}]{2019A&A...631A..72L}
{Lin}, Y., {Csengeri}, T., {Wyrowski}, F., {et~al.} 2019, \aap, 631, A72, \dodoi{10.1051/0004-6361/201935410}

\bibitem[{{Liu} {et~al.}(2022{\natexlab{a}}){Liu}, {Qiu}, \& {Zhang}}]{2022ApJ...925...30L}
{Liu}, J., {Qiu}, K., \& {Zhang}, Q. 2022{\natexlab{a}}, \apj, 925, 30, \dodoi{10.3847/1538-4357/ac3911}

\bibitem[{{Liu} {et~al.}(2021){Liu}, {Zhang}, {Commer{\c{c}}on}, {Valdivia}, {Maury}, \& {Qiu}}]{2021ApJ...919...79L}
{Liu}, J., {Zhang}, Q., {Commer{\c{c}}on}, B., {et~al.} 2021, \apj, 919, 79, \dodoi{10.3847/1538-4357/ac0cec}

\bibitem[{{Liu} {et~al.}(2022{\natexlab{b}}){Liu}, {Zhang}, \& {Qiu}}]{2022FrASS...9.3556L}
{Liu}, J., {Zhang}, Q., \& {Qiu}, K. 2022{\natexlab{b}}, Frontiers in Astronomy and Space Sciences, 9, 943556, \dodoi{10.3389/fspas.2022.943556}

\bibitem[{{Liu} {et~al.}(2020){Liu}, {Zhang}, {Qiu}, {Liu}, {Pillai}, {Girart}, {Li}, \& {Wang}}]{2020ApJ...895..142L}
{Liu}, J., {Zhang}, Q., {Qiu}, K., {et~al.} 2020, \apj, 895, 142, \dodoi{10.3847/1538-4357/ab9087}

\bibitem[{{Machida} \& {Basu}(2020)}]{2020MNRAS.494..827M}
{Machida}, M.~N., \& {Basu}, S. 2020, \mnras, 494, 827, \dodoi{10.1093/mnras/staa672}

\bibitem[{{Maury} {et~al.}(2018){Maury}, {Girart}, {Zhang}, {Hennebelle}, {Keto}, {Rao}, {Lai}, {Ohashi}, \& {Galametz}}]{2018MNRAS.477.2760M}
{Maury}, A.~J., {Girart}, J.~M., {Zhang}, Q., {et~al.} 2018, \mnras, 477, 2760, \dodoi{10.1093/mnras/sty574}

\bibitem[{{McKee} \& {Ostriker}(2007)}]{2007ARA&A..45..565M}
{McKee}, C.~F., \& {Ostriker}, E.~C. 2007, \araa, 45, 565, \dodoi{10.1146/annurev.astro.45.051806.110602}

\bibitem[{{McMullin} {et~al.}(2007){McMullin}, {Waters}, {Schiebel}, {Young}, \& {Golap}}]{2007ASPC..376..127M}
{McMullin}, J.~P., {Waters}, B., {Schiebel}, D., {Young}, W., \& {Golap}, K. 2007, in Astronomical Society of the Pacific Conference Series, Vol. 376, Astronomical Data Analysis Software and Systems XVI, ed. R.~A. {Shaw}, F.~{Hill}, \& D.~J. {Bell}, 127

\bibitem[{{M{\"o}ller} {et~al.}(2017){M{\"o}ller}, {Endres}, \& {Schilke}}]{2017A&A...598A...7M}
{M{\"o}ller}, T., {Endres}, C., \& {Schilke}, P. 2017, \aap, 598, A7, \dodoi{10.1051/0004-6361/201527203}

\bibitem[{{Mottram} {et~al.}(2011){Mottram}, {Hoare}, {Davies}, {Lumsden}, {Oudmaijer}, {Urquhart}, {Moore}, {Cooper}, \& {Stead}}]{2011ApJ...730L..33M}
{Mottram}, J.~C., {Hoare}, M.~G., {Davies}, B., {et~al.} 2011, \apjl, 730, L33, \dodoi{10.1088/2041-8205/730/2/L33}

\bibitem[{{Mouschovias} \& {Ciolek}(1999)}]{1999ASIC..540..305M}
{Mouschovias}, T.~C., \& {Ciolek}, G.~E. 1999, in NATO Advanced Study Institute (ASI) Series C, Vol. 540, The Origin of Stars and Planetary Systems, ed. C.~J. {Lada} \& N.~D. {Kylafis}, 305

\bibitem[{{Mouschovias} \& {Paleologou}(1979)}]{1979ApJ...230..204M}
{Mouschovias}, T.~C., \& {Paleologou}, E.~V. 1979, \apj, 230, 204, \dodoi{10.1086/157077}

\bibitem[{{Olguin} {et~al.}(2022){Olguin}, {Sanhueza}, {Ginsburg}, {Chen}, {Zhang}, {Li}, {Lu}, \& {Sakai}}]{2022ApJ...929...68O}
{Olguin}, F.~A., {Sanhueza}, P., {Ginsburg}, A., {et~al.} 2022, \apj, 929, 68, \dodoi{10.3847/1538-4357/ac5bd8}

\bibitem[{{Olguin} {et~al.}(2021){Olguin}, {Sanhueza}, {Guzm{\'a}n}, {Lu}, {Saigo}, {Zhang}, {Silva}, {Chen}, {Li}, {Ohashi}, {Nakamura}, {Sakai}, \& {Wu}}]{2021ApJ...909..199O}
{Olguin}, F.~A., {Sanhueza}, P., {Guzm{\'a}n}, A.~E., {et~al.} 2021, \apj, 909, 199, \dodoi{10.3847/1538-4357/abde3f}

\bibitem[{{Olguin} {et~al.}(2023){Olguin}, {Sanhueza}, {Chen}, {Lu}, {Oya}, {Zhang}, {Ginsburg}, {Taniguchi}, {Li}, {Morii}, {Sakai}, \& {Nakamura}}]{2023ApJ...959L..31O}
{Olguin}, F.~A., {Sanhueza}, P., {Chen}, H.-R.~V., {et~al.} 2023, \apjl, 959, L31, \dodoi{10.3847/2041-8213/ad1100}

\bibitem[{{Ossenkopf} \& {Henning}(1994)}]{1994A&A...291..943O}
{Ossenkopf}, V., \& {Henning}, T. 1994, \aap, 291, 943

\bibitem[{{Ostriker} {et~al.}(2001){Ostriker}, {Stone}, \& {Gammie}}]{2001ApJ...546..980O}
{Ostriker}, E.~C., {Stone}, J.~M., \& {Gammie}, C.~F. 2001, \apj, 546, 980, \dodoi{10.1086/318290}

\bibitem[{{Padovani} \& {Galli}(2011)}]{2011A&A...530A.109P}
{Padovani}, M., \& {Galli}, D. 2011, \aap, 530, A109, \dodoi{10.1051/0004-6361/201116853}

\bibitem[{{Padovani} {et~al.}(2013){Padovani}, {Hennebelle}, \& {Galli}}]{2013A&A...560A.114P}
{Padovani}, M., {Hennebelle}, P., \& {Galli}, D. 2013, \aap, 560, A114, \dodoi{10.1051/0004-6361/201322407}

\bibitem[{{Padovani} {et~al.}(2012){Padovani}, {Brinch}, {Girart}, {J{\o}rgensen}, {Frau}, {Hennebelle}, {Kuiper}, {Vlemmings}, {Bertoldi}, {Hogerheijde}, {Juhasz}, \& {Schaaf}}]{2012A&A...543A..16P}
{Padovani}, M., {Brinch}, C., {Girart}, J.~M., {et~al.} 2012, \aap, 543, A16, \dodoi{10.1051/0004-6361/201219028}

\bibitem[{{Palau} {et~al.}(2021){Palau}, {Zhang}, {Girart}, {Liu}, {Rao}, {Koch}, {Estalella}, {Chen}, {Liu}, {Qiu}, {Li}, {Zapata}, {Bontemps}, {Ho}, {Beuther}, {Ching}, {Shinnaga}, \& {Ahmadi}}]{2021ApJ...912..159P}
{Palau}, A., {Zhang}, Q., {Girart}, J.~M., {et~al.} 2021, \apj, 912, 159, \dodoi{10.3847/1538-4357/abee1e}

\bibitem[{{Pineda} {et~al.}(2015){Pineda}, {Offner}, {Parker}, {Arce}, {Goodman}, {Caselli}, {Fuller}, {Bourke}, \& {Corder}}]{2015Natur.518..213P}
{Pineda}, J.~E., {Offner}, S. S.~R., {Parker}, R.~J., {et~al.} 2015, \nat, 518, 213, \dodoi{10.1038/nature14166}

\bibitem[{{Price} \& {Bate}(2007)}]{2007MNRAS.377...77P}
{Price}, D.~J., \& {Bate}, M.~R. 2007, \mnras, 377, 77, \dodoi{10.1111/j.1365-2966.2007.11621.x}

\bibitem[{{Qiu} {et~al.}(2014){Qiu}, {Zhang}, {Menten}, {Liu}, {Tang}, \& {Girart}}]{2014ApJ...794L..18Q}
{Qiu}, K., {Zhang}, Q., {Menten}, K.~M., {et~al.} 2014, \apjl, 794, L18, \dodoi{10.1088/2041-8205/794/1/L18}

\bibitem[{{Rao} {et~al.}(2009){Rao}, {Girart}, {Marrone}, {Lai}, \& {Schnee}}]{2009ApJ...707..921R}
{Rao}, R., {Girart}, J.~M., {Marrone}, D.~P., {Lai}, S.-P., \& {Schnee}, S. 2009, \apj, 707, 921, \dodoi{10.1088/0004-637X/707/2/921}

\bibitem[{{Rosolowsky} {et~al.}(2008){Rosolowsky}, {Pineda}, {Kauffmann}, \& {Goodman}}]{2008ApJ...679.1338R}
{Rosolowsky}, E.~W., {Pineda}, J.~E., {Kauffmann}, J., \& {Goodman}, A.~A. 2008, \apj, 679, 1338, \dodoi{10.1086/587685}

\bibitem[{{Sanhueza} {et~al.}(2021){Sanhueza}, {Girart}, {Padovani}, {Galli}, {Hull}, {Zhang}, {Cortes}, {Stephens}, {Fern{\'a}ndez-L{\'o}pez}, {Jackson}, {Frau}, {Kock}, {Wu}, {Zapata}, {Olguin}, {Lu}, {Silva}, {Tang}, {Sakai}, {Guzm{\'a}n}, {Tatematsu}, {Nakamura}, \& {Chen}}]{2021ApJ...915L..10S}
{Sanhueza}, P., {Girart}, J.~M., {Padovani}, M., {et~al.} 2021, \apjl, 915, L10, \dodoi{10.3847/2041-8213/ac081c}

\bibitem[{{Soler} \& {Hennebelle}(2017)}]{2017A&A...607A...2S}
{Soler}, J.~D., \& {Hennebelle}, P. 2017, \aap, 607, A2, \dodoi{10.1051/0004-6361/201731049}

\bibitem[{{Stephens} {et~al.}(2013){Stephens}, {Looney}, {Kwon}, {Hull}, {Plambeck}, {Crutcher}, {Chapman}, {Novak}, {Davidson}, {Vaillancourt}, {Shinnaga}, \& {Matthews}}]{2013ApJ...769L..15S}
{Stephens}, I.~W., {Looney}, L.~W., {Kwon}, W., {et~al.} 2013, \apjl, 769, L15, \dodoi{10.1088/2041-8205/769/1/L15}

\bibitem[{{Tang} {et~al.}(2009){Tang}, {Ho}, {Koch}, {Girart}, {Lai}, \& {Rao}}]{2009ApJ...700..251T}
{Tang}, Y.-W., {Ho}, P. T.~P., {Koch}, P.~M., {et~al.} 2009, \apj, 700, 251, \dodoi{10.1088/0004-637X/700/1/251}

\bibitem[{{Taniguchi} {et~al.}(2023){Taniguchi}, {Sanhueza}, {Olguin}, {Gorai}, {Das}, {Nakamura}, {Saito}, {Zhang}, {Lu}, {Li}, \& {Chen}}]{2023ApJ...950...57T}
{Taniguchi}, K., {Sanhueza}, P., {Olguin}, F.~A., {et~al.} 2023, \apj, 950, 57, \dodoi{10.3847/1538-4357/acca1d}

\bibitem[{{Vaillancourt}(2006)}]{2006PASP..118.1340V}
{Vaillancourt}, J.~E. 2006, \pasp, 118, 1340, \dodoi{10.1086/507472}

\end{thebibliography}
\bibliographystyle{aasjournal}



\end{document}